# Analytically approximate solution to the VLE problem with the SRK equation of state

Hongqin Liu*

Integrated High Performance Computing Branch, Shared Services Canada, Montreal, QC, Canada

Abstract

Since a transcendental equation is involved in vapor liquid equilibrium (VLE) calculations with a cubic equation of state (EoS), any exact solution has to be carried out numerically with an iterative approach [1,2]. This causes significant wastes of repetitive human efforts and computing resources. Based on a recent study [3] on the Maxwell construction [4] and the van der Waals EoS [5], here we propose a procedure for developing analytically approximate solutions to the VLE calculation with the Soave-Redlich-Kwong (SRK) EoS [6] for the entire coexistence curve. This procedure can be applied to any cubic EoS and thus opens a new area for the EoS study. For industrial applications, a simple databank can be built containing only the coefficients of a newly defined function and other thermodynamic properties will be obtained with analytical forms. For each system there is only a one-time effort, and therefore, the wastes caused by the repetitive efforts can be avoided. By the way, we also show that for exact solutions, the VLE problem with any cubic EoS can be reduced to solving a transcendental equation with one unknown, which can significantly simplify the methods currently employed [2,7].

*Emails: hongqin.liu@canada.ca; hqliu2000@gmail.com.

## Introduction

Vapor-liquid equilibrium (VLE) calculation plays a central role in thermodynamics and process simulations [1,7]. With an equation of state (EoS), the chemical potential equilibrium condition, or the Maxwell constructions [4], generates a transcendental function with logarithm terms. As a result, exact solutions have to be carried out numerically by some iterative methods. The pressure equilibrium condition and the transcendental equation are usually solved together, which is not a simple task and becomes a research topic by itself [2]. Worst of all, the efforts made by one individual will have to be repeated by another even for the same system at the same conditions, unless numerical data are published, which is not practical. The costs of computing resources are significant as well. As reported [7], the phase equilibrium computation is the most expensive step in chemical process simulations.

Some efforts have been made for developing analytical solutions, but mostly in the neighbourhood of the critical point [8,9]. The method aimed at the diameter [10,11] of the coexistence curve seems pointing to a right direction, but it is limited by the fact that both saturated liquid and vapor volumes (densities) are complex functions of temperature, which makes even empirical correlations difficult. In practical applications, engineers have to use empirical correlations for those most useful properties. A well-known example is the Antoine equation [13] for vapor pressure (the equilibrium pressure of a pure system) correlations. Others properties, including heat capacity, saturated densities etc., have to be correlated separately [13]. Theoretically, if an EoS works good enough for a given system, the saturated volumes of the liquid and vapor phases will provide all the information needed for any thermodynamic properties along entire the coexistence.

Moreover, for some applications where derivative properties, such as the vapor pressure vs temperature, numerical solutions face difficulties. But with analytical solutions, this task usually becomes minor, if not trivial. Consequently, analytical solution to the VLE problem is the ultimate goal for engineers and researchers.

When a cubic EoS is applied to the VLE calculation, subjected to the Maxwell construction, which is equivalent to the combination of the pressure and chemical potential equilibrium conditions, two real roots are for saturated volumes of liquid and vapor phases, respectively, and the third "unphysical" root is the Maxwell crossover or the M-line [3]. It is found that the M-line is a smooth function of temperature, in contrast to the volumes of saturated liquid and vapor phases. By taking advantages of special features of the M-line, we will propose a procedure for developing analytically approximate solutions to the VLE problem for a given cubic EoS. We demonstrate the procedure with the Soave-Redlich-Kwong (SRK) EoS [6] by using the information at the critical point from Singley, Burns and Misovich (1997) [9].

## Exact solution to the VLE problem by a cubic EoS

Before targeting an analytical solution, here we first show that the VLE problem with a cubic EoS can always be reduced to solving a transcendental equation with one unknown. We use the SRK EoS as the example to prove the statement and it is applicable to any cubic EoS. The SRK EoS reads [6]:

$$P = \frac{RT}{v-b} - \frac{a(T_r)}{v(v+b)} \quad (1)$$

where, $P$ is pressure, $T$, temperature, $v$, molar volume, $R$, the gas constant, $a(T_r)$ is a function of temperature (see the Appendix) and $b$ is a constant. For the SRK EoS, due to the adoption of a third parameter (species-dependent), the acentric factor, $\omega$, all the calculations have to be based on individual substance cases. From the chemical potential equilibrium condition, we have:

$$\frac{v_G}{v_G-b} - \frac{v_L}{v_L-b} - \frac{\theta b}{(v_G+b)} + \frac{\theta b}{(v_L+b)} = \\ ln\frac{v_G-b}{v_L-b} + \theta\left(ln\frac{v_G+b}{v_G} - ln\frac{v_L+b}{v_L}\right) \quad (2)$$

where $\theta = a(T_r)/(RTb)$, $v_L$ and $v_G$ are molar volumes for liquid and vapor phases, respectively. For the exact solutions, we will generate a quadratic equation first. Throughout of this work, we only consider the VLE case where temperature is given. Applying the pressure equilibrium condition to liquid and vapor phases (see the Appendix, Eq.(S1)), we get:

$$D_G v^3 - v^2 - (D_G b + 1 - \theta)bv - \theta b^2 = 0 \quad (3)$$

where the liquid volume, $v_L$, has been replaced by a dummy variable, $v$, and

$$D_G = \frac{1}{v_G-b} - \frac{\theta b}{v_G(v_G+b)} \quad (4)$$

Eq.(3) is a cubic function. It is easy to prove that at temperature, $T \geq T_c$, there is only one root and as $T < T_c$ there are always three real (> 0) roots, and we denote them as $\alpha, \beta, \gamma$. According to the root-coefficient relations for a generic function, $px^3 + qx^2 + rx + s = 0$, we know that:

$$\alpha + \beta + \gamma = -\frac{q}{p} = \frac{1}{D_G} \quad (5)$$

$$\alpha\beta\gamma = -\frac{s}{p} = \frac{\theta b^2}{D_G} \qquad (6)$$

$$\alpha\beta + \beta\gamma + \alpha\gamma = \frac{r}{p} = -\frac{D_G b + 1 - \theta}{D_G} \qquad (7)$$

Now we assume that $\alpha, \beta$ represent two volumes: $v_L$, $v_M$, respectively, where $v_L$ is the liquid molar volume and $v_M$, the unphysical intermediate volume named as the volume of M phase. From Eq.(5)-(7), we can easily prove that $\gamma = v_G$, namely the vapor volume. Now we can construct a quadratic equation that has the same two roots ($v_L$, $v_M$) as those from Eq.(3):

$$v^2 + uv + w = 0 \qquad (8)$$

where

$$u = -\frac{1}{D_G} + v_G \qquad (9)$$

$$w = \frac{\theta b^2}{D_G v_G} \qquad (10)$$

Then we have the solution for liquid volume:

$$v_L = \frac{-u - \sqrt{u^2 - 4w}}{2} = f(v_G) \qquad (11)$$

where $f(v_G)$ is a given function of $v_G$. In Eq.(2), replacing $v_L$ with Eq.(11), we have a transcendental equation with only one unknown, $v_G$, and this ends our proof for previous statement. ∎

With traditional method [3], the pressure and chemical equilibrium conditions are solved at the time and high quality initial guesses are demanded. The new method significantly simplifies the solution process. It is found that the transcendental equation, Eq.(2) and (11), is stable given a reasonable initial volume. It can be easily solved with the Excel Solver. All the exact solutions used in this work are obtained this way.

## The Maxwell construction and the M-line

A full discussion on the Maxwell construction (known as the equal-area rule, EAR) and the crossover is presented in [3] with the van der Waals EoS as an example. **Figure 1** illustrates the Maxwell construction and related quantities to help our presentations.

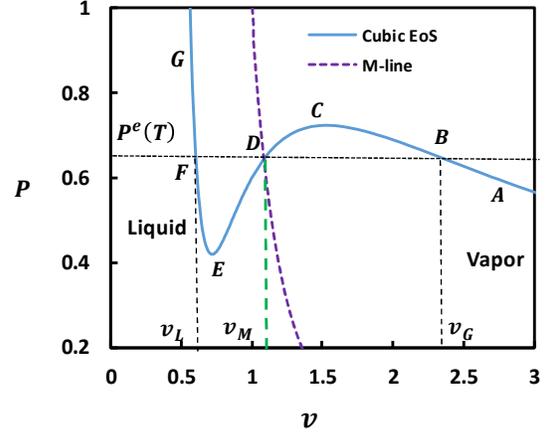

**Figure 1**. The Maxwell construction and the M-line. At equilibrium condition: area FEDF = area DBCD.

The Maxwell can be written analytically as:

$$P^e = \frac{1}{v_G - v_L}\int_{v_L}^{v_G} P dv_r \qquad (12)$$

where $P^e$ is the equilibrium pressure at a given temperature. By re-arrangement, we have:

$$\int_{v_L}^{v_M}(P^e - P)dv_r = \int_{v_M}^{v_G}(P - p^e)dv_r \qquad (13)$$

which is the EQR. Eq.(13) means that area FEDF = area DBCD. Therefore, the EQR generates an intermediate volume, $v_M$. It should be mentioned that the Maxwell construction applies to any EoS, not necessarily a cubic one. As a cubic EoS is adopted, the value of $v_M$ is the "unphysical" third root since there is only one intersection between the pressure curve $P(v)$ and the equilibrium pressure, $P^e$. From Figure 1, we also see that:

$$P(v_L) = P(v_M) = P(v_G) = p^e \qquad (14)$$

The state at $v_M$ represents an unstable state, since $\mu(v_M) \neq \mu(v_L)$ or $\mu(v_G)$, where $\mu$ is the chemical potential. This means that a thermodynamic property at this state is not a equilibrium property. Since we are only interested in the volume itself, which turns out to be physically meaningful, the treatments below are physically sound. The trajectory of the intermediate volume, $v_M$, is called the Maxwell crossover or the M-line (Figure 1). The M-line can be easily calculated by the SRK EoS. In fact, the "unphysical" root of Eq.(8) is the intermediate volume as function of $v_G$:

$$v_M = \frac{-u + \sqrt{u^2 - 4w}}{2} \qquad (15)$$

Therefore, as discussed above, the solution to the transcendental equation, Eq.(2), will give both $v_L$ and $v_M$ at a given temperature by Eq.(11) and (15), respectively, as $v_G$ is known. Similarly, we can rewrite Eq.(3) as

$$D_M v^3 - v^2 - (D_M b + 1 - \theta)bv - \theta b^2 = 0 \quad (16)$$

where $D_M$ is defined the same way as $D_G$ by Eq.(4) with $v_G$ being replaced by $v_M$. Therefore we have:

$$v_{L|G} = \frac{-u \mp \sqrt{u^2 - 4w}}{2} \quad (17)$$

where the subscript, $L|G$, correspond to $\mp$, respectively. Eq.(17) provides $v_L$ and $v_G$ as $v_M$ is known. From Eq.(11) and (17) we have the diameter of the coexistence curve[10]:

$$d_\sigma = \frac{\rho_L + \rho_G}{2} = \frac{1 - v_M D_M}{2 D_M} \frac{D_M v_M}{\theta b^2} = \frac{v_M(1 - v_M D_M)}{2\theta b^2} \quad (18)$$

Or we can simply get $v_M$ as a function of $d_\sigma$. This result shows that the "unphysical root" does have a physical meaning. Now we define an entropic function as following:

$$\mathcal{S} = ln(\tilde{v}_M - 1) = ln\left(\frac{v_M}{b} - 1\right) \quad (19)$$

Figure 2 plots the function, $\mathcal{S}$, for three phases for ethane, where M-phase is represented by the M-line. This figure shows a remarkable comparison between the three phases: M-line behaves much smoother than the other two. Therefore, we use the M-line as the base for developing our analytical solutions.

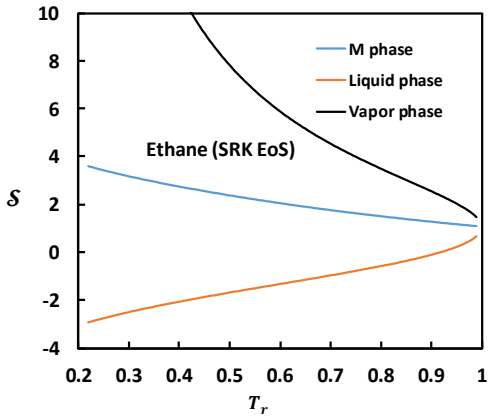

**Figure 2**. Plots of the function $\mathcal{S} = ln(\tilde{v}_M - 1)$ for different phases (ethane). All curves are calculated with exact solutions, Eq.(2), (11) and (15).

## The analytical solution to the VLE problem with the SRK EoS

For the purpose of obtaining high accurate functions, the entire temperature range is divided into two regions: $0 < T_r \leq T_{r0}$ and $T_{r0} < T_r \leq 1$. The characteristic temperature, $T_{r0}$, is determined empirically, such that best results for both regions can be obtained. After some tests, the final definition is:

$$T_{r0} = \frac{2}{5}\left(\frac{T_C}{T_{CAr}}\right)^{\frac{1}{5}} \quad (20)$$

where $T_{CAr}$ is the critical temperature of argon, and the empirical constant, 2/5, is only for the substances considered in this work (mainly hydrocarbons). One may choose other values with small deviations from Eq.(20), but other coefficients discussed below will have to be changed accordingly. In the low temperature range, $T_r \leq T_{r0}$, from the pressure and chemical potential equilibrium conditions (see the Appendix), we can get analytically approximate solutions:

$$\frac{v_L}{b} \approx \frac{1}{2}\left(\theta - 1 - \sqrt{1 - 6\theta + \theta^2}\right) \quad (21)$$

$$v_G \approx e(v_L - b)\left(\frac{v_L + b}{v_L}\right)^\theta \quad (22)$$

where $e$ is the natural constant. In the temperature range, $T_{r0} < T_r \leq 1$, we use a polynomial function:

$$\mathcal{S} = ln(\tilde{v}_M - 1) = \sum_{i=0}^{5} C_i T_r^i \quad (23)$$

where $\tilde{v}_M = v_M/b$. Eq.(23) has 6 coefficients, from which we have $v_M = b(1 + e^\mathcal{S})$. At low temperature end point, $T_r = T_{r0}$, we impose the following constraints generated from chemical potential condition:

$$\left.\frac{d^n \mu_G}{dT^n}\right|_0 = \left.\frac{d^n \mu_L}{dT^n}\right|_0, n = 0, 1, 2 \quad (24)$$

3 coefficients can be determined by Eq.(24), where $v_L$ and $v_G$ are calculated with Eq.(21) and (22), respectively. Given SRK chemical potential and the volume expressions discussed above, it is straightforward to use Eq.(24) (see the Appendix for details) for determining three coefficients.

For other 3 coefficients, we use the information at the critical point, which requires some elaboration. Singley, Burns and Misovich (1997) [9] have derived a series expansion at the critical point up to 10th order. The reduced saturated densities of liquid and vapor phases are given by the following, respectively:

$$\rho_{rL} = 1 + B_1(1 - T_r)^{\frac{1}{2}} + B_2(1 - T_r) + B_3(1 - T_r)^{\frac{3}{2}} + B_4(1 - T_r)^2 + B_5(1 - T_r)^{\frac{5}{2}} + B_6(1 - T_r)^3 + \cdots \quad (25)$$

$$\rho_{rG} = 1 - B_1(1-T_r)^{\frac{1}{2}} + B_2(1-T_r) - B_3(1-T_r)^{\frac{3}{2}} + B_4(1-T_r)^2 - B_5(1-T_r)^{\frac{5}{2}} + B_6(1-T_r)^3 - \cdots \quad (26)$$

We define the following functions:

$$X = \rho_{rL} + \rho_{rG} = 2 + 2B_2(1-T_r) + 2B_4(1-T_r)^2 + 2B_6(1-T_r)^3 + \cdots \quad (27)$$

$$y = \rho_{rL}\rho_{rG} = 1 + (2B_2 - B_1^2)(1-T_r) + (2B_4 - 2B_1B_3 + B_2B_2)(1-T_r)^2 +$$
$$(2B_6 - 2B_1B_5 + 2B_2B_4 - B_3B_3)(1-T_r)^3 + (-2B_3B_5 + 2B_2B_6 + B_4B_4 + \cdots)(1-T_r)^4 -$$
$$B_5B_5(1-T_r)^5 + B_4B_6(1-T_r)^5 + B_6B_4(1-T_r)^5 \ldots \quad (28)$$

Where the coefficients, $B_2$, $B_4$ etc. are given functions of the acentric factor [9]. Then we have:

$$v_L + v_G = \frac{v_C}{\rho_{rL}} + \frac{v_C}{\rho_{rG}} = \frac{(\rho_{rL} + \rho_{rG})v_C}{\rho_{rL}\rho_{rG}} \equiv \frac{v_C X}{y} \quad (29)$$

Eq.(17) leads to:

$$v_L + v_G = -u = \frac{1}{D_M} - v_M = \frac{1 - v_M D_M}{D_M} \quad (30)$$

And

$$v_L v_G = w = \frac{\theta b^2}{D_M v_M} \quad (31)$$

where

$$D_M = \frac{1}{v_M - b} - \frac{\theta b}{v_M(v_M + b)} \quad (32)$$

At the critical point, we have:

$$D_C = \frac{1}{v_C - b} + \theta_C\left(\frac{1}{v_C + b} - \frac{1}{v_C}\right) \quad (33)$$

$$\left.\frac{dD_M}{dT}\right|_C = \left(\frac{1}{v_C + b} - \frac{1}{v_C}\right)\left.\frac{d\theta}{dT}\right|_C \quad (34)$$

$$\left.\frac{d^2 D_M}{dT^2}\right|_C = 2\left(\frac{1}{v_C^2} - \frac{1}{(v_C + b)^2}\right)\left.\frac{d\theta}{dT}\right|_C \left.\frac{dv_M}{dT}\right|_C +$$
$$\left(\frac{1}{v_C + b} - \frac{1}{v_C}\right)\left.\frac{d^2\theta}{dT^2}\right|_C \quad (35)$$

Where we have used the critical conditions:

$$\left(\frac{\partial D_M}{\partial v}\right)_T = -\frac{1}{(v_C - b)^2} + \theta_C\left(\frac{1}{v_C^2} - \frac{1}{(v_C + b)^2}\right) = 0 \quad (36)$$

$$\left(\frac{\partial^2 D_M}{\partial v^2}\right)_T = \frac{2}{(v_C - b)^3} + \theta_C\left(-\frac{2}{v_C^3} + \frac{2}{(v_C + b)^3}\right) = 0 \quad (37)$$

Finally:

$$\left.\frac{dv_M}{dT}\right|_C = -\frac{1}{D_C^2}\left.\frac{dD_M}{dT}\right|_C - \frac{1}{T_C}\left(\left.\frac{dv_L}{dT_r}\right|_C + \left.\frac{dv_G}{dT_r}\right|_C\right) \quad (38)$$

$$\left.\frac{d^2 v_M}{dT^2}\right|_C = -\frac{1}{D_C^2}\left.\frac{d^2 D_M}{dT^2}\right|_C + \frac{2}{D_C^3}\left(\left.\frac{dD_M}{dT}\right|_C\right)^2 -$$
$$\frac{1}{T_C^2}\left(\left.\frac{d^2 v_L}{dT_r^2}\right|_C + \left.\frac{d^2 v_G}{dT_r^2}\right|_C\right) \quad (39)$$

Therefore, at the critical point we have three quantities, which are functions of $B_i$ ($i$=1,2…):

$$\left.\frac{d^n v_M}{dT^n}\right|_C, n = 0, 1, 2 \quad (40)$$

The missing details about calculating the above derivatives are provided in the Appendix. Eq.(40) will be used to determine 3 coefficients in Eq.(23) with linear functions. For the SRK EoS, $S_C = 1.046384$. In all the calculations, the critical volume is calculated from $v_C = Z_C R T_C / P_C$.

In total we have 6 equations and 3 are non-linear functions, Eq.(24). The calculations have been carried in the Excel Solver. The derivatives required for the chemical potential constraints are derived from Eq.(17) (Appendix).

In summary, in the low temperature range, $T_r \leq T_{r0}$, Eq.(21) and (22) will be used to calculate the volumes. In the high temperature range, $T_{r0} < T_r \leq 1$, Eq.(23) and (17) will used to calculate the saturated volumes. Finally in the entire temperature range, the equilibrium pressure is calculated by

$$P^e = \frac{RT}{v_G - v_L}\ln\frac{v_G - b}{v_L - b} - \frac{a(T)}{b(v_G - v_L)}\ln\frac{v_G(v_L + b)}{v_L(v_G + b)} \quad (41)$$

Which can be obtained from Eq.(1) and (12). From the analytical expressions for the volumes and pressure, we can derive all other thermodynamic properties along the entire coexistence curve analytically as they are functions of $v_G$ and $v_L$. Here we provide a few of them. Enthalpy is given by:

$$H = H^* + RT(Z - 1) - \frac{a}{b}\left(1 - \frac{Ta'}{a}\right)\ln\frac{v + b}{v} \quad (42)$$

where $a' = da/dT$ (Appendix). Then we have the latent heat, $\Delta H_v$, from the enthalpy:

$$\frac{\Delta H_v}{RT} = Z_G - Z_L - \frac{a - a'T}{RTb}\ln\frac{v_L(v_G + b)}{v_G(v_L + b)} \quad (43)$$

The isobaric heat capacity is given by:

$$\frac{C_P}{R} = \frac{C_P^{id}}{R} - 1 + \frac{a''(T)}{bR} \ln\frac{v+b}{v} +$$
$$\frac{v^2(v+b)^2 - 2R^{-1}a'(T)v(v^2-b^2) + R^{-2}a'^2(T)(v-b)^2}{v^2(v+b)^2 - a(T)(RT)^{-1}(2v+b)(v-b)^2} \quad (44)$$

where the heat capacity of the ideal gas, $C_P^{id}$, is from Ref [12]. $a''(T)$ is provided in Appendix.

Another advantage of using the analytical solution is for derivative properties. For example, the well-know Clapeyron equation correlates the latent heat with the vapor pressure derivative [1,8]:

$$\frac{\Delta H_v}{P_C v_C} = \frac{RT}{P_C v_C} \frac{v_G - v_L}{R} \frac{dP^e}{dT} \quad (45)$$

A thermodynamic consistency testing is the comparison between the results from Eq.(43) and (45). By numerical solution, using Eq.(45) would be difficult since numerical differentiation is required. With analytical solution, applying Eq.(45) is a minor effort. $\frac{dP^e}{dT}$ can be obtained from Eq.(41) and the volume derivatives from Eq.(17) (see Appendix for details).

For demonstration purpose, in this work we totally considered 8 substances for vapor pressure predictions and more details are presented for argon and ethane. The critical properties and other constants are listed in Table 1. The coefficients of Eq.(23) obtained for those substances are listed in Table 2. Other data used for the calculations are listed in Table S1.

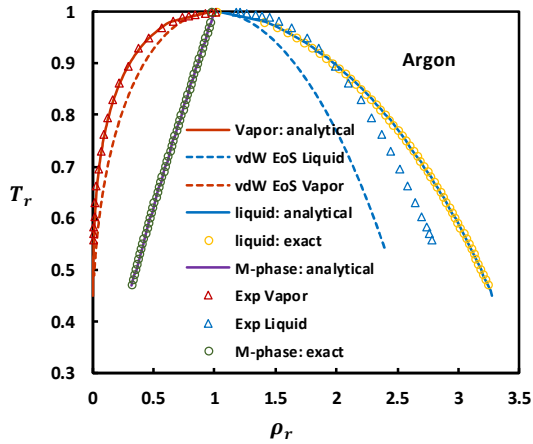

**Figure 3**. Calculation results with the SRK EoS for densities of Argon. Open triangles are experimental data [13]; open circles are exact solutions; solid lines are from analytical solutions; dashed lines are from vdW EoS, which obviously cannot serve quantitative prediction purpose.

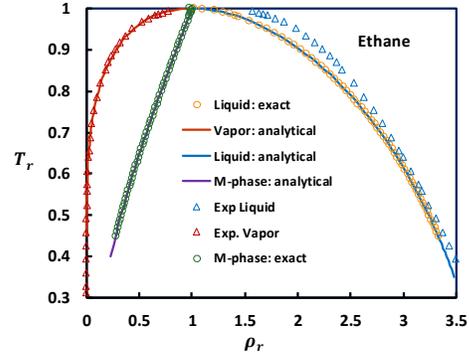

**Figure 4**. Calculation results with the SRK EoS for densities of Ethane. Open triangles are experimental data [14]; open circles are exact solutions; solid lines are from analytical solutions.

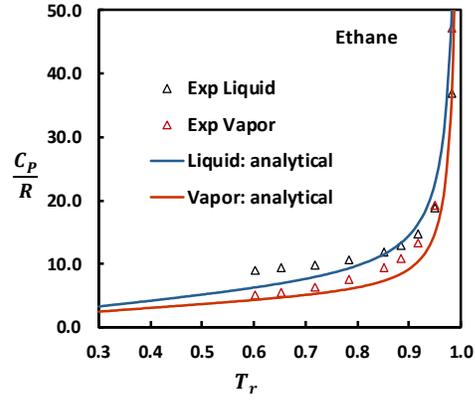

**Figure 5**. Isobaric heat capacity calculated by the analytical solutions, Eq.(44). The differences between the exact and analytical solutions are negligible compared to those between the experimental data [15] and predictions.

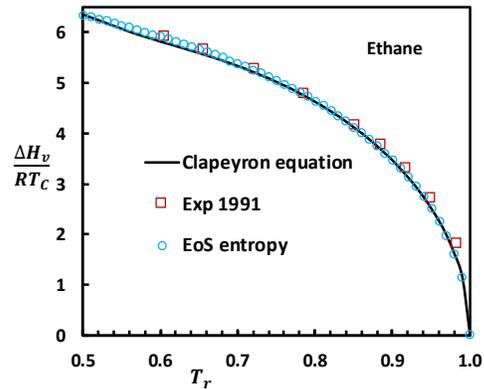

**Figure 6**. Thermodynamic constancy testing for the latent heat. The results from the Clapeyron equation, Eq.(45), are from the analytical solution . The differences between the exact and analytical solutions are again negligible. The experimental data are from ref [15].

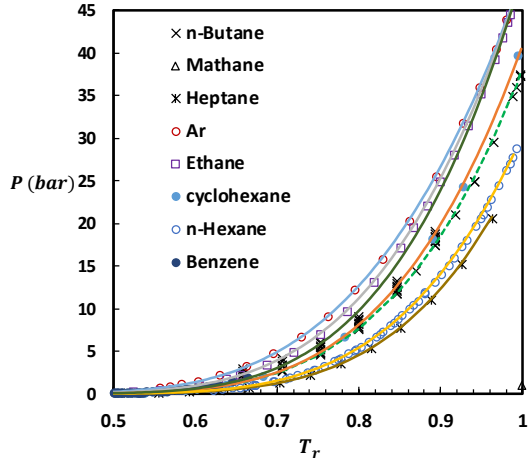

**Figure 7**. The equilibrium pressure prediction results by the SRK EoS for all the systems considered in this work. The solid curves are calculation results, from Eq.(41) with analytical volume expressions, and points are experimental data. The data sources: Ar, [13]; Methane, [16]; Ethane, [14]; n-Butane, [17]; cyclohexane, [19,20]; n-Hexane, [18]; n-Heptane, [21]; Benzene, [22].

Figure 3 and Figure 4 illustrate the densities calculated by the analytical solution and the exact solutions for Argon and Ethane, respectively. Compared with the prediction errors against the experimental data, the discrepancies between the analytical solutions and exact solutions are all negligible. Figure 5 depicts the isobaric heat capacities calculated by the analytical solutions, Eq.(44), (17) and (23), vs experimental data. Figure 6 depicts the thermodynamic consistency testing with the SRK EoS, calculated by Eq.(43) and (45), respectively. The agreements between the two methods are excellent. The application of the Clapeyron equation requires $dP^e/dT$, which is calculated with Eq.(41) and the derivatives of saturated volumes are calculated from the analytical solutions, Eq.(17) and (23). The details are provided in the Appendix.

Figure 7 presents the equilibrium pressure calculated by Eq.(41) with analytical expressions for the volumes. The agreements between the analytical solutions and exact solutions are above 2-orders of magnitude higher than those when compared with the experimental data, therefore satisfy the needs of all practical applications.

Finally, as an example, Table 3 lists a comparison between the exact and analytical solutions for ethane. The worst case is at $T_r = 0.6$ and the low temperature end at $T_r = 0.46$. The table shows that for liquid volume and the equilibrium pressure the agreements are excellent. For vapor volume the agreement is expectable.

Table 1 Critical properties and constants used in the calculations

|  | $T_C, K$ | $P_C, bar$ | $v_C, l/mol$ | $\omega$ | $b$ | $a_c$ |
|---|---|---|---|---|---|---|
| Argon | 150.8 | 48.7 | 0.0858 | 0.001 | 0.022306 | 1.3799 |
| Methane | 190.4 | 46 | 0.1147 | 0.011 | 0.029817 | 2.3289 |
| Ethane | 305.4 | 48.8 | 0.1734 | 0.099 | 0.045082 | 5.6480 |
| n-Butane | 425.2 | 38 | 0.3101 | 0.199 | 0.080605 | 14.0597 |
| n-Hexane | 507.5 | 30.1 | 0.4673 | 0.299 | 0.121457 | 25.2860 |
| Cyclohexane | 553.8 | 40.7 | 0.3771 | 0.212 | 0.098019 | 22.2682 |
| n-Heptane | 540.3 | 27.4 | 0.5465 | 0.349 | 0.142049 | 31.4842 |
| Benzene | 562.1 | 48.9 | 0.3186 | 0.212 | 0.082805 | 19.0938 |

$v_C = Z_C R T_C / P_C = R T_C / 3 P_C$. The critical constants are from ref [12].

Table 2 Coefficients of Eq.(23) and AAD for pressure

| Substance | $T_{r0}$ | $C_0$ | $C_1$ | $C_2$ | $C_3$ | $C_4$ | $C_5$ | AAD% |
|---|---|---|---|---|---|---|---|---|
| Argon | 0.40000 | 4.722378 | -6.806245 | 4.570508 | -1.460235 | -0.123006 | 0.142984 | 0.0042 |
| Methane | 0.41910 | 4.662219 | -6.239253 | 2.922949 | 0.763598 | -1.575823 | 0.512695 | 0.0060 |
| Ethane | 0.46063 | 4.719780 | -5.846706 | 1.998728 | 1.310195 | -1.586006 | 0.450395 | 0.0041 |
| n-Butane | 0.49215 | 4.781632 | -5.445759 | 1.037975 | 1.942411 | -1.694007 | 0.424133 | 0.0043 |
| n-Pentane | 0.50205 | 4.536968 | -3.398230 | -4.296358 | 8.498408 | -5.674155 | 1.379751 | 0.0013 |
| cyclohexane | 0.51886 | 4.935708 | -6.329727 | 3.253134 | -0.841432 | 0.017716 | 0.010986 | 0.0037 |
| n-Hexane | 0.50988 | 4.501827 | -2.875797 | -5.475596 | 9.607763 | -6.181930 | 1.470117 | 0.0008 |
| n-Heptane | 0.51631 | 4.504431 | -2.621153 | -5.923348 | 9.760085 | -6.085093 | 1.411463 | 0.0014 |
| Benzene | 0.52041 | 4.543005 | -3.727193 | -3.484854 | 7.684354 | -5.266515 | 1.297588 | 0.0008 |

$AAD\% = \frac{100}{N_p} \sum_{i=1}^{N_p} \frac{abs(P^e_{analytic} - P^e_{exact})}{P^e_{exact}}$, where the total number of data points, $N_p = 70$, which equally divide a temperature range: $0.3 \leq T_r \leq 1$.

**Table 3** A comparison between the results from the exact and analytic solutions for ethane

| $T_r$ | $v_L$ (exact) | $v_G$ (exact) | $P^e$ (exact) | $v_L$ (analytic) | $v_G$ (analytic) | $P^e$ (analytic) |
|---|---|---|---|---|---|---|
| 0.46 | 0.0523601 | 308.11 | 0.0378294 | 0.0523603 | 309.47 | 0.0378291 |
| 0.6 | 0.0571321 | 15.983 | 0.927126 | 0.0571319 | 15.704 | 0.926985 |

## Conclusions and discussions

This article presents a procedure for deriving explicit solutions to the VLE problem with a cubic EoS over the entire temperature range. To author's knowledge, no such work has ever been reported before. This work will change the way the VLE calculations are carried out with a cubic EoS. The repetitive calculations are replaced with one-time effort, namely finding the coefficients of Eq.(23) (or alike) by solving few non-linear equations. In addition, for the exact numerical solutions to the VLE problem using a cubic EoS, we show that the pressure and chemical potential equilibrium conditions can always be reduced into a single transcendental function with one unknown. This makes the exact solution easier than solving two equations at the same time.

This work suggests some new areas in cubic EoS studies. For demonstration purpose, we only discussed the SRK EoS. There are dozens of cubic EoS's which are widely used in various applications[1,23-25]. Many of them are more accurate than the original SRK EoS [6]. It is expected that the method developed here can be applied to any cubic EoS. First of all, the series expansions at the critical point need to be carried out for the EoS interested. The work of Singley, Burns and Misovich (1997) [9] provides such an example.

One of the objectives of using current procedure is that the analytical equations possess prediction power, which means that all properties can be predicted by using the properties at two "end" points only, namely the critical point and the characteristic low temperature point, $T_{r0}$. It is found that in a narrow "middle temperature (density) range the predicted vapor volumes are less accurate, sometimes with the average absolute deviation (AAD) of 1+%, when compared with the exact solutions. The equilibrium pressure, Eq.(41), and liquid density prediction are with much higher accuracies (see Table 2). For industrial applications, the accuracies for predicted properties are good enough.

The accuracies can be improved by selecting few points (exact solutions) in the middle range and re-fitting the coefficients. After all, this is a one-time effort only and one can take other approaches. If the high order series expansions are not available at the critical point, one could only take some points from the exact solutions to fit the coefficients. The functional forms of (23) are also a mater of choices, as mentioned.

Finally, for a thermodynamic-property databank establishment used in practical applications, the data needed are only the coefficients of the M-line, Eq.(23) in this case. The rest will be some analytical equations and the prediction results will be thermodynamically consistent.

## Acknowledgement

The author is grateful to Dr. Misovich for providing his research information and publications.

## Appendix

**The SRK EoS and related relations**

The pressure equilibrium condition for three phases:

$$P^e = \frac{RT}{v_L - b} - \frac{a(T_r)}{v_L(v_L + b)} = \frac{RT}{v_G - b} - \frac{a(T_r)}{v_G(v_G + b)} = \frac{RT}{v_M - b} - \frac{a(T_r)}{v_M(v_M + b)} \quad (S1)$$

where $a(T_r) = a_C \alpha(T_r)$ [6] and

$$\alpha(T_r) = \left[1 + f(\omega)\left(1 - T_r^{\frac{1}{2}}\right)\right]^2 \quad (S2)$$

The function for the acentric factor is given by:

$$f(\omega) = 0.480 + 1.574\omega - 0.176\omega^2 \quad (S3)$$

and:

$$a_C = 0.42747 \frac{R^2 T_C^2}{P_C}, \qquad b = 0.08664 \frac{RT_C}{P_C} \quad (S4)$$

$$\theta = \frac{a(T)}{RTb}, \qquad \theta_C = \frac{a_c}{RT_C b} = 4.933864 = \Omega \quad (S5)$$

From Eq.(S5), we get:

$$\frac{d\theta}{dT} = \frac{a'(T)}{RTb} - \frac{a(T)}{RT^2 b}, \qquad \frac{d^2\theta}{dT^2} = \frac{a''(T)}{RTb} - \frac{2a'(T)}{RT^2 b} - \frac{2a(T)}{RT^3 b} \quad (S6)$$

$$a' = \frac{da}{T_C dT_r} = -\frac{a_c f(\omega)}{T_C \sqrt{T_r}}\left[1 + f(\omega)\left(1 - T_r^{\frac{1}{2}}\right)\right] \quad (S7)$$

$$a'' = \frac{a_c f(\omega)}{2T_C^2 T_r^{\frac{3}{2}}} \left[ 1 + f(\omega)\left(1 - T_r^{\frac{1}{2}}\right) + \frac{f(\omega)}{\sqrt{T_r}} \right] \tag{S8}$$

From Eq.(41), we have:

$$\begin{aligned}
\frac{dP^e}{dT} &= \frac{R}{v_G - v_L} \ln\frac{v_G - b}{v_L - b} - \frac{a'(T)}{b(v_G - v_L)}\left[\ln\frac{v_G}{v_L} - \ln\frac{v_G + b}{v_L + b}\right] \\
&+ \frac{RT}{(v_G - v_L)^2}\left[\theta\left(\ln\frac{v_G}{v_L} - \ln\frac{v_G + b}{v_L + b}\right) - \ln\frac{v_G - b}{v_L - b}\right]\left(\frac{dv_G}{dT} - \frac{dv_L}{dT}\right) \\
&+ \frac{RT}{v_G - v_L}\left[\left(\frac{1}{v_G - b} - \frac{\theta}{v_G} + \frac{\theta}{v_G + b}\right)\frac{dv_G}{dT} - \left(\frac{1}{v_L - b} - \frac{\theta}{v_L} + \frac{\theta}{v_L + b}\right)\frac{dv_L}{dT}\right]
\end{aligned} \tag{S9}$$

**Low temperature approximations**

For self completeness, few equations appeared in the main test may reappear here. From the SRK EoS:

$$\frac{P}{RT} = \frac{1}{v_L - b} - \frac{\theta b}{v_L(v_L + b)} = D_G = \frac{1}{v_G - b} - \frac{\theta b}{v_G(v_G + b)} \tag{S10}$$

where

$$D_G = \frac{1}{v_G - b} - \frac{\theta b}{v_G(v_G + b)} \tag{S10a}$$

From Eq.(9) we have:

$$D_G v_L^3 - v_L^2 - (D_G b + 1 - \theta)bv_L - \theta b^2 = 0 \tag{S11}$$

Low temperature, $T_r \leq T_{r0}$, $v_G \gg b$, $v_G \gg b\theta$. In fact, since $\theta > 1$ the low temperature condition can be simply defined as $v_G \gg b\theta$. From Eq.(10a) we have, $D_G \approx 1/v_G$, which leads to ($D_G v_L^3 \sim 0$, $D_G b \sim 0$), then Eq.(S1) is simplified to:

$$v_L^2 + (1 - \theta)bv_L + \theta b^2 = 0 \tag{S12}$$

Therefore:

$$v_L \approx \frac{b}{2}\left(\theta - 1 - \sqrt{1 - 6\theta + \theta^2}\right) \tag{S13}$$

This equation provides high accuracy for $v_L$, as $T_r \leq T_{r0}$. However, equilibrium pressure cannot be calculated from it since the chemical potential equilibrium condition has not been involved. Therefore, for $v_G$, the chemical equilibrium condition, Eq.(2), has to be imposed. At low T, $v_G \gg b$, $v_G \gg b\theta$, $Z^G \to 1$ (ideal gas), $Z^L \to 0$, finally we have:

$$\ln v_G = 1 + \theta \ln\frac{v_L + b}{v_L} + \ln(v_L - b) \tag{S14}$$

$v_L$ and $v_G$ calculated from the approximate equations agree with strict solution with high precisions. The differences between the equilibrium pressures from the strict and approximate solutions at low temperature are negligible (with ADD<0.01%). For example, for ethane, $b = 0.045083, , l/mol$, $T_{r0} = 0.46063$, at $T_r = 0.46$, $b\theta = 0.70098$, strict $v_G = 308.11, l/mol$, therefore $v_G \gg b\theta$ holds solidly. The equilibrium pressure (for all temperature range) can be calculated with Eq.(41). By the way, from Eq.(S13) and (S14) we get the derivatives:

$$\frac{dv_L}{dT} \approx \frac{b}{2}\left[1 + \frac{3 - \theta}{\sqrt{1 - 6\theta + \theta^2}}\right]\frac{d\theta}{dT} \tag{S15}$$

$$\frac{1}{v_G}\frac{dv_G}{dT} = \ln\frac{v_L + b}{v_L}\frac{d\theta}{dT} + \left(\frac{\theta}{v_L + b} - \frac{\theta}{v_L} + \frac{1}{v_L - b}\right)\frac{dv_L}{dT} \tag{S16}$$

$$\frac{d^2 v_L}{dT^2} = \frac{b}{2}\left[1 + \frac{3 - \theta}{\sqrt{1 - 6\theta + \theta^2}}\right]\frac{d^2\theta}{dT^2} + \frac{4b}{(1 - 6\theta + \theta^2)^{\frac{3}{2}}}\left(\frac{d\theta}{dT}\right)^2 \tag{S17}$$

$$\frac{1}{v_G}\frac{d^2 v_G}{dT^2} = \frac{1}{v_G^2}\left(\frac{dv_G}{dT}\right)^2 + \ln\frac{v_L + b}{v_L}\frac{d^2\theta}{dT^2} - \frac{2b}{v_L(v_L + b)}\frac{d\theta}{dT}\frac{dv_L}{dT} +$$
$$\left[\frac{\theta}{v_L^2} - \frac{\theta}{(v_L + b)^2} - \frac{1}{(v_L - b)^2}\right]\left(\frac{dv_L}{dT}\right)^2 + \left(\frac{\theta}{v_L + b} - \frac{\theta}{v_L} + \frac{1}{v_L - b}\right)\frac{d^2 v_L}{dT^2} \quad (S18)$$

The above equations can be used in the low temperature range as needed. For example for calculation of the pressure derivative, Eq.(S9), the first derivatives are required.

**Derivatives for coefficient evaluations for Eq.(23)**

As show in the main test, we define:

$$D_M = \frac{1}{v_M - b} - \frac{\theta b}{v_M(v_M + b)} \quad (S19)$$

$$u = -\frac{1}{D_M} + v_M \quad (S20)$$

$$w = \frac{\theta b^2}{D_M v_M} \quad (S21)$$

Then:

$$v_{L|G} = \frac{-u \mp \sqrt{u^2 - 4w}}{2} \quad (S22)$$

From the above, we have the derivatives:

$$\frac{dD_M}{dT} = \left[-\frac{1}{(v_M - b)^2} + \frac{\theta b}{v_M(v_M + b)}\left(\frac{1}{v_M} + \frac{1}{v_M + b}\right)\right]\frac{dv_M}{dT} - \frac{b}{v_M(v_M + b)}\frac{d\theta}{dT} \quad (S23)$$

$$\frac{du}{dT} = \frac{1}{D_M^2}\frac{dD_M}{dT} + \frac{dv_M}{dT} \quad (S24)$$

$$\frac{dw}{dT} = -\frac{\theta b^2}{D_M v_M}\left(\frac{1}{D_M}\frac{dD_M}{dT} + \frac{1}{v_M}\frac{dv_M}{dT}\right) + \frac{b^2}{D_M v_M}\frac{d\theta}{dT} \quad (S25)$$

$$\frac{dv_G}{dT} = -\frac{1}{2}\left(1 - \frac{u}{\sqrt{u^2 - 4w}}\right)\frac{du}{dT} - \frac{1}{\sqrt{u^2 - 4w}}\frac{dw}{dT} \quad (S26)$$

$$\frac{dv_L}{dT} = -\frac{1}{2}\left(1 + \frac{u}{\sqrt{u^2 - 4w}}\right)\frac{du}{dT} + \frac{1}{\sqrt{u^2 - 4w}}\frac{dw}{dT} \quad (S27)$$

$$\frac{d^2 D_M}{dT^2} = \left[-\frac{1}{(v_M - b)^2} + \frac{\theta b}{v_M^2(v_M + b)} + \frac{\theta b}{v_M(v_M + b)^2}\right]\frac{d^2 v_M}{dT^2}$$
$$+ \left[\frac{2}{(v_M - b)^3} - \frac{2\theta b}{v_M^3(v_M + b)} - \frac{2\theta b}{v_M^2(v_M + b)^2} - \frac{2\theta b}{v_M(v_M + b)^3}\right]\left(\frac{dv_M}{dT}\right)^2$$
$$+ 2\left[\frac{b}{v_M^2(v_M + b)} + \frac{b}{v_M(v_M + b)^2}\right]\frac{dv_M}{dT}\frac{d\theta}{dT} - \frac{b}{v_M(v_M + b)}\frac{d^2\theta}{dT^2} \quad (S28)$$

$$\frac{d^2 u}{dT^2} = \frac{1}{D_M^2}\frac{d^2 D_M}{dT^2} - \frac{2}{D_M^3}\left(\frac{dD_M}{dT}\right)^2 + \frac{d^2 v_M}{dT^2} \quad (S29)$$

$$\frac{d^2 w}{dT^2} = g_1 - \frac{2b^2}{D_M v_M}\left[\frac{1}{D_M}\frac{dD_M}{dT} + \frac{1}{v_M}\frac{dv_M}{dT}\right]\frac{d\theta}{dT} - \frac{b^2 \theta}{D_M v_M}\left[\frac{1}{D_M}\frac{d^2 D_M}{dT^2} + \frac{1}{v_M}\frac{d^2 v_M}{dT^2} - \frac{1}{\theta}\frac{d^2\theta}{dT^2}\right] \quad (S30)$$

$$g_1 = \frac{2b^2\theta}{D_M v_M}\left[\frac{1}{D_M^2}\left(\frac{dD_M}{dT}\right)^2 + \frac{1}{D_M v_M}\frac{dv_M}{dT}\frac{dD_M}{dT} + \frac{1}{v_M^2}\left(\frac{dv_M}{dT}\right)^2\right] \quad (S31)$$

From Eq.(S26) and (S27), we have:

$$\frac{d^2 v_L}{dT^2} = -U - U1 - U2 + \frac{1}{\Delta}\frac{d^2 w}{dT^2} \quad (S32)$$

$$\frac{d^2 v_G}{dT^2} = -U + U1 + U2 - \frac{1}{\Delta}\frac{d^2 w}{dT^2} \tag{S33}$$

where:

$$\Delta = \sqrt{u^2 - 4w} \tag{S34a}$$

$$U = \frac{1}{2}\left(1 - \frac{u}{\Delta}\right)\frac{d^2 u}{dT^2} \tag{S34b}$$

$$U1 = \frac{1}{2\Delta}\frac{du}{dT}\left[\frac{du}{dT} - \frac{u}{u^2 - 4w}\left(u\frac{du}{dT} - 2\frac{dw}{dT}\right)\right] \tag{S34c}$$

$$U2 = \frac{1}{(u^2 - 4w)^{\frac{3}{2}}}\left(u\frac{du}{dT} - 2\frac{dw}{dT}\right)\frac{dw}{dT} \tag{S34d}$$

The above equations will be used when applying the chemical potential constraints at the low temperature end, $T_{r0}$ and at the critical point. The terms involving $v_M$ are given by Eq.(23) as functions of the unknown coefficients:

$$v_M = b(1 + e^{\mathcal{S}}) \tag{S35}$$

**Determining the coefficients of Eq.(23)**

$$\mathcal{S} = \ln(\tilde{v} - 1) = C_0 + C_1 T_r + C_2 T_r^2 + C_3 T_r^3 + C_4 T_r^4 + C_5 T_r^5 \tag{S36}$$

$$\frac{d\mathcal{S}}{dT_r} = C_1 + 2C_2 T_r + 3C_3 T_r^2 + 4C_4 T_r^3 + 5C_5 T_r^4 \tag{S37}$$

$$\frac{d^2\mathcal{S}}{dT_r^2} = 2C_2 + 6C_3 T_r + 12C_4 T_r^2 + 20C_5 T_r^3 \tag{S38}$$

At the critical point:

$$\mathcal{S}_C = C_0 + C_1 + C_2 + C_3 + C_4 + C_5 \tag{S39}$$

$$\mathcal{S}'_C = \frac{d\mathcal{S}}{dT_r} = C_1 + 2C_2 + 3C_3 + 4C_4 + 5C_5 \tag{S40}$$

$$\mathcal{S}''_C = \frac{d^2\mathcal{S}}{dT_r^2} = 2C_2 + 6C_3 + 12C_4 + 20C_5 \tag{S41}$$

At the low temperature, $T_{r0}$, we have:

$$\mathcal{S}_0 = \ln(\tilde{v}_{M0} - 1) = C_0 + C_1 T_{r0} + C_2 T_{r0}^2 + C_3 T_{r0}^3 + C_4 T_{r0}^4 + C_5 T_{r0}^5 \tag{S42}$$

$$\mathcal{S}'_0 = \left.\frac{d\mathcal{S}}{dT_r}\right|_0 = C_1 + 2C_2 T_{r0} + 3C_3 T_{r0}^2 + 4C_4 T_{r0}^3 + 5C_5 T_{r0}^4 \tag{S43}$$

$$\mathcal{S}''_0 = \left.\frac{d^2\mathcal{S}}{dT_r^2}\right|_0 = 2C_2 + 6C_3 T_{r0} + 12C_4 T_{r0}^2 + 20C_5 T_{r0}^3 \tag{S44}$$

$$\left.\frac{dv_M}{dT}\right|_0 = \frac{v_{M0} - b}{T_C}\left.\frac{d\mathcal{S}}{dT_r}\right|_0 = \mathcal{S}'_0 \frac{(v_{M0} - b)}{T_C} \tag{S45}$$

$$\left.\frac{d^2 v_M}{dT^2}\right|_0 = \frac{(v_{M0} - b)}{T_C^2}(\mathcal{S}''_0 + \mathcal{S}'^2_0) \tag{S46}$$

**Chemical potential constraints at $T_{r0}$ for determining the coefficients, $C_3, C_4, C_5$.**

Zeroth derivative (condition 1):

$$Z^G - \ln(v_G - b) - \theta \ln\left(\frac{v_G + b}{v_G}\right) = Z^L - \ln(v_L - b) - \theta \ln\left(\frac{v_L + b}{v_L}\right) \tag{S47}$$

Using the definition of the compressibility, we have Eq.(2). It's first derivative (condition 2):

$$\left[\frac{\theta}{v_G} - \frac{v_G}{(v_G - b)^2} - \frac{\theta v_G}{(v_G + b)^2}\right]\frac{dv_G}{dT} + \left[\ln\frac{v_G}{v_G + b} - \frac{b}{(v_G + b)}\right]\frac{d\theta}{dT}$$
$$= \left[\frac{\theta}{v_L} - \frac{v_L}{(v_L - b)^2} - \frac{\theta v_L}{(v_L + b)^2}\right]\frac{dv_L}{dT} + \left[\ln\frac{v_L}{v_L + b} - \frac{b}{(v_L + b)}\right]\frac{d\theta}{dT}$$

(S48)

The second derivatives (condition 3):

$$\left[\frac{\theta}{v_G} - \frac{v_G}{(v_G - b)^2} - \frac{\theta v_G}{(v_G + b)^2}\right]\frac{d^2 v_G}{dT^2} - \left[\frac{\theta}{v_L} - \frac{v_L}{(v_L - b)^2} - \frac{\theta v_L}{(v_L + b)^2}\right]\frac{d^2 v_L}{dT^2}$$
$$+ \left[-\frac{\theta}{v_G^2} + \frac{v_G + b}{(v_G - b)^3} + \frac{\theta(v_G - b)}{(v_G + b)^3}\right]\left(\frac{dv_G}{dT}\right)^2 + 2\left[\frac{1}{v_G} - \frac{v_G}{(v_G + b)^2}\right]\frac{dv_G}{dT}\frac{d\theta}{dT}$$
$$- \left[-\frac{\theta}{v_L^2} + \frac{v_L + b}{(v_L - b)^3} + \frac{\theta(v_L - b)}{(v_L + b)^3}\right]\left(\frac{dv_L}{dT}\right)^2 - 2\left[\frac{1}{v_L} - \frac{v_L}{(v_L + b)^2}\right]\frac{d\theta}{dT}\frac{dv_L}{dT}$$
$$= \left(\frac{b}{v_G + b} - \frac{b}{v_L + b} + \ln\frac{v_G + b}{v_L + b} - \ln\frac{v_G}{v_L}\right)\frac{d^2\theta}{dT^2}$$

(S49)

The derivatives required in Eq.(S48) and (S49) are given by Eq.(S43)-Eq.(S46), which are functions of $C_3, C_4, C_5$. Now we need to find $S_C$, $S_C'$ and $S_C''$ for other three coefficients (linear functions) of Eq.(S36).

**Calculations of $S_C'$ and $S_C''$**

For the critical volume, we calculate it from other two critical constants:

$$v_C = \frac{Z_C R T_C}{P_C}$$

(S50)

Then we have:

$$\frac{v_C}{b} = \frac{Z_C}{0.08664} = 3.84734$$

(S51)

Therefore, the following constant is universal:

$$S_C = \ln\left(\frac{v_C}{b} - 1\right) = 1.046384$$

(S52)

From definition, Eq.(29), we have

$$\left.\frac{dv_L}{dT_r}\right|_C + \left.\frac{dv_G}{dT_r}\right|_C = v_C\left(\left.\frac{dX}{dT_r}\right|_C - 2\left.\frac{dY}{dT_r}\right|_C\right)$$

(S53)

$$\left.\frac{d^2 v_L}{dT_r^2}\right|_C + \left.\frac{d^2 v_G}{dT_r^2}\right|_C = v_C\left[\left.\frac{d^2 X}{dT_r^2}\right|_C - 2\left.\frac{d^2 Y}{dT_r^2}\right|_C - 2\left.\frac{dX}{dT_r}\right|_C\left.\frac{dY}{dT_r}\right|_C + 4\left(\left.\frac{dY}{dT_r}\right|_C\right)^2\right]$$

(S54)

where

$$\left.\frac{dv_L}{dT}\right|_C = \frac{1}{T_C}\left.\frac{dv_L}{dT_r}\right|_C, \qquad \left.\frac{dv_G}{dT}\right|_C = \frac{1}{T_C}\left.\frac{dv_G}{dT_r}\right|_C$$

(S55)

$$\left.\frac{d^2 v_L}{dT^2}\right|_C = \frac{1}{T_C^2}\left.\frac{d^2 v_L}{dT_r^2}\right|_C, \qquad \left.\frac{d^2 v_G}{dT^2}\right|_C = \frac{1}{T_C^2}\left.\frac{d^2 v_G}{dT_r^2}\right|_C$$

(S56)

Before moving on, we need to calculate $X, Y$ and their derivatives at the critical point. The work of Singley, Burns and Misovich (1997) [9] provides the key to this. For SRK EoS the authors used series expansion at the critical point for pressure and density, Eq.(25) and (26). At the critical point, from Eq.(27) and (28) we have:

$$X_C = 2, \qquad \left.\frac{dX}{dT_r}\right|_C = -2B_2, \qquad \left.\frac{d^2 X}{dT_r^2}\right|_C = 4B_4$$

(S57)

$$y_C = 1, \quad \left.\frac{dy}{dT_r}\right|_C = -(2B_2 - B_1^2), \quad \left.\frac{d^2y}{dT_r^2}\right|_C = 2(2B_4 - 2B_1 B_3 + B_2 B_2) \tag{S58}$$

Where the constants have been provided by Singley et al. [9], which are functions of $f = f(\omega)$, Eq.(S3):

$$B_1 = 2.25992\sqrt{1+f}, \quad B_2 = 0.98283(1+f) \tag{S59}$$

$$B_3 = -\sqrt{1+f}(0.33227 + 1.17974f) \tag{S60}$$

$$B_4 = -0.05345 - 0.84402f - 0.79057f^2 \tag{S61}$$

Finally, we can calculate $S'_C$ and $S''_C$.

$$S'_C = \left.\frac{dS}{dT_r}\right|_C = \frac{T_C}{v_C - b}\left.\frac{dv_M}{dT}\right|_C = -\frac{T_C}{v_C - b}\left[\frac{1}{D_C^2}\left.\frac{dD_M}{dT}\right|_C + \frac{1}{T_C}\left(\left.\frac{dv_L}{dT_r}\right|_C + \left.\frac{dv_G}{dT_r}\right|_C\right)\right] \tag{S62}$$

$$S''_C = \left.\frac{d^2S}{dT_r^2}\right|_C = \frac{T_C^2}{v_C - b}\left[\left.\frac{d^2v_M}{dT^2}\right|_C - \frac{1}{v_C - b}\left(\left.\frac{dv_M}{dT}\right|_C\right)^2\right] \tag{S63}$$

By using Eq.(S52) and the last two equations back to Eq.(S39)-(S41) to replace $C_0, C_1$ and $C_2$, we have three unknowns, $C_3, C_4, C_5$. By solving the three equations, Eq.(S47)-(S49), we find all the coefficients for Eq.(23). In this work the Excel solver was used for the calculations. Table S1 lists calculated values for $B_i, i = 1, \ldots 4, S'_C$ and $S''_C$ for the substances discussed in this work.

Table S1 Constants at the critical point

|             | $B_1$   | $B_2$   | $B_3$    | $B_4$    | $S'_C$   | $S''_C$  |
|-------------|---------|---------|----------|----------|----------|----------|
| Argon       | 2.75077 | 1.45614 | -1.09597 | -0.64325 | -1.82304 | 1.76321  |
| Methane     | 2.76533 | 1.47158 | -1.12446 | -0.66868 | -1.84238 | 1.77150  |
| Ethane      | 2.88890 | 1.60604 | -1.38103 | -0.90652 | -2.01072 | 1.83444  |
| n-Butane    | 3.02040 | 1.75559 | -1.68380 | -1.20580 | -2.19794 | 1.88498  |
| n-Hexane    | 3.14356 | 1.90167 | -1.99637 | -1.53349 | -2.38083 | 1.91457  |
| Cyclohexane | 3.03686 | 1.77477 | -1.72392 | -1.24684 | -2.22196 | 1.88998  |
| n-Heptane   | 3.20231 | 1.97341 | -2.15571 | -1.70722 | -2.47065 | 1.92194  |
| Benzene     | 3.03686 | 1.77477 | -1.72392 | -1.24684 | -2.22196 | 1.88998  |